# AXISYMMETRIC CIRCUMSTELLAR INTERACTION IN SUPERNOVAE

John M. Blondin

Department of Physics, North Carolina State University, Raleigh NC 27695-8202;

john_blondin@ncsu.edu

Peter Lundqvist

Stockholm Observatory, S-133 36 Saltsjöbaden, Sweden; peter@astro.su.se

and

Roger A. Chevalier

Department of Astronomy, University of Virginia, P.O. Box 3818, Charlottesville, VA 22903; rac5x@virginia.edu

# ABSTRACT

Multiwavelength observations of Type II supernovae have shown evidence for the interaction of supernovae with the dense slow winds from the red supergiant progenitor stars. Observations of planetary nebulae and the nebula around SN 1987A show that the slow winds from extended stars frequently have an axisymmetric structure with a high density in the equatorial plane. We have carried out numerical calculations of the interaction of a supernova with such an axisymmetric density distribution. For small values of the angular density gradient at the pole, the asymmetry in the interaction shell is greater than, but close to, that expected from purely radial motion. If the angular density gradient is above a moderate value, the flow qualitatively changes and a protrusion emerges along the axis. For a power-law supernova density profile, the flow approaches a self-similar state in which the protrusion length is $2-4$ times the radius of the main shell. The critical density gradient is larger for steeper density profiles of the ejecta. Most of our calculations are axisymmetric, but we have carried out a 3-dimensional calculation to show that the protrusion is not a numerical artifact along the symmetry axis. For typical supernova parameters, the protrusions take $\gtrsim$ several years to develop. The appearance of the shell with protrusions is similar to that observed in VLBI radio images of the remnant 41.9 +58 in M82 and, possibly, of SN 1986J. We also considered the possibility of asymmetric ejecta and found that it had a relatively small effect on the asymmetry of the interaction region.

*Subject headings:* hydrodynamics – shock waves – supernova remnants – supernovae:general

# 1. INTRODUCTION

Previous models of supernovae (SNe) of Types Ib, Ic and II interacting with circumstellar gas have relied on the assumption that the circumstellar medium (CSM), created by the stellar wind of the progenitor, has a spherically symmetric density distribution (e.g., Chevalier 1982; Chevalier, Blondin, & Emmering 1992; Chevalier & Fransson 1994). For most SNe this has proven to be a good assumption in order to explain the observations (Chevalier 1984; Chevalier & Fransson 1994), but there is now a growing consensus that at least for some SNe, non-spherical symmetry may be important.

The ring around SN 1987A (e.g., Jakobsen et al. 1991) and the bipolar nebula connected to this ring (Wampler et al. 1991; Wang & Wampler 1992; Burrows et al. 1995; Plait et al. 1995) probably constitute the clearest evidence for non-spherical symmetry. To explain the observed structure Luo & McCray (1990), Wang & Mazzali (1991), Blondin & Lundqvist (1993), Martin & Arnett (1995) and Chevalier & Dwarkadas (1995) used the assumption that the progenitor had a spherically symmetric blue supergiant (BSG) wind which was interacting with a previously lost non-spherically symmetric red supergiant (RSG) wind. The hydrodynamical models by Blondin & Lundqvist (1993) show that if the structure was formed in this way, a high ratio ($\gtrsim 20$) of equatorial to polar mass loss during the RSG stage is necessary. The interacting winds model has gained support from imaging observations by Crotts, Kunkel, & Heathcote (1995); the observations by Crotts et al. show both the equatorial ring and a structure connected to this which is similar to that in the model.

Other models for the formation of the CSM around SN 1987A invoke a non-spherically symmetric BSG wind interacting with a spherically symmetric RSG wind (Blondin 1994; Chevalier & Luo 1994), the possibility that the ring is protostellar (McCray & Lin 1994), or that the structure of the CSM is a result of the wind from the progenitor interacting with that from a binary companion (Podsiadlowski, Fabian, & Stevens 1990). Regardless of which model turns out to be closest to the real situation, SN 1987A will start interacting with its CSM in a non-spherically symmetric fashion no later than when it collides with the ring (e.g., Luo, McCray, & Slavin 1994). At least in the model with an asymmetric RSG wind it is clear that the circumstellar interaction would have been quite different from spherical already from start, had the SN exploded while still a RSG.

VLBI (Very Long Baseline Interferometry) radio observations of SNe are an especially powerful technique for investigating asymmetric interaction because it is possible to image emission from the interaction region. Marcaide et al. (1995a,b) have produced a series of images of SN 1993J over the age range 6 to 18 months and found that the image is remarkably symmetric, although there are brightness variations around the shell. However, the other two SNe for which there is VLBI imaging, SN 1986J in NGC 891 and 41.9 +58 in M82, show significant asymmetry. SN 1986J, which is thought to have occurred in 1982, shows three protrusions from a shell (Bartel et al. 1991). The remnant 41.9 +58, which is thought to have occurred in the mid-1950's, shows a shell with two protrusions on opposite sides (Bartel et al. 1987; Wilkinson & de Bruyn 1990). The images are poorly resolved, but the outer radii of the protrusions are $\sim 2-3$ times the shell radius. Chevalier & Blondin (1995) investigated the possibility that the protrusions are the result of Rayleigh-Taylor instabilities in the decelerating thin shell of ejecta which has radiatively cooled. In fact, the unstable fingers did not distort the outer shock front, so the influence of inhomogeneous density structure was indicated for the protrusions.

Optical line emission can also be an indicator of asymmetric or inhomogeneous structure. Chevalier & Fransson (1994) suggested that in order to explain the strong narrow lines from SNe such as SN 1986J or SN 1988Z, a deviation from spherical interaction may be needed. Chugai & Danziger (1994) have explored the same idea and find that the CSM of SN 1988Z could either be clumpy, or have a high density in the equatorial plane. Recently,

Cumming et al. (1996) argue along the same lines for SN 1994W.

Also SN remnants show some evidence for being relics of SNe interacting with non-spherically symmetric surroundings. Igumenshchev, Tutukov, & Shustov (1992) find, in an analysis similar to that in this paper, that the asphericity present for $\sim 30\%$ of remnants observed using the Einstein Observatory (Seward 1990), can be explained if the progenitor had stronger mass loss in the equatorial plane than along the poles. In an attempt to explain the evolution of the optical structure of the synchrotron emission from the Crab Nebula, Fesen, Martin, & Shull (1992) suggested that also the Crab progenitor may have experienced asymmetric mass loss shortly before the explosion.

The fact that some SNe are expected to interact with an asymmetric CSM is not surprising from the structure observed around some RSGs and red giants. In particular, $\mu$ Cep (e.g., Mauron & Querci 1990) shows evidence for asymmetry, and in the study by Plez & Lambert (1994) on the four red giants R Aql, V Hya, g Her and R Leo, it was found that all four had a highly asymmetrical distribution of K I $\lambda 7699$ emission. R Aql, and other Mira variables, also show evidence for asymmetry in the VLA study by Bowers, Johnston, & De Vegt (1989), along with the supergiant S Per. In addition, Trammell, Dinerstein, & Goodrich (1994) find in their study of post-AGB stars that a majority of the stars displayed intrinsic polarization, presumably due to an aspherical distribution of circumstellar matter. These findings may support the currently most popular model for the shaping of planetary nebulae (PNe), in which the fast wind from the central star interacts with a previously lost cylindrically symmetric red giant wind (Kwok, Purton, & Fitzgerald 1978; Kahn & West 1985), i.e., very similar to the interacting winds model for the CSM of SN 1987A. If this model is correct, non-spherically symmetric red giant (and presumably also RSG) winds should be common. In fact, Zuckerman & Aller (1986) find that out of 108 PNe, $\sim 50\%$ displayed bipolar symmetry.

There is thus both observational and theoretical support for studying the interaction of SNe with a non-spherically symmetric CSM. It is of particular interest to study the stability of spherically symmetric circumstellar interaction against asymmetry, and how different polar density distributions of the circumstellar matter affect the SN/wind interaction. We present our numerical model for the SN/CSM interaction in §2. In §3 we discuss the results of our numerical simulations, including the dependence on the asymmetry of the CSM, the dependence on the ejecta density profile, the effect of enforcing axisymmetry, and the effect of asymmetric ejecta. The implications of these results concerning observed SNe are discussed in §4.

## 2. MODEL

The numerical model used to study the asphericity of the SN/wind interaction is similar to that used in Chevalier et al. (1992): Two-dimensional hydrodynamic simulations of a self-similar driven wave (SSDW) using the numerical hydrodynamics code VH-1, a multidimensional code based on the PPM algorithm of Colella & Woodward (1984). A spherical SSDW was initialized on a two-dimensional numerical grid spanning one quadrant of a sphere, i.e., assuming axisymmetry about the polar axis and reflection symmetry about the equator. The grid was expanded in the radial direction as the SSDW evolved, so that the interaction could be followed for many expansion times. The gas is assumed to be adiabatic, with a ratio of specific heats $\gamma = 5/3$. The SSDW is initialized with an outer shock velocity of $10^9$ cm s$^{-1}$ at a radius of $10^{13}$ cm. Unless otherwise specified, all of the models shown here have an ejecta density profile described by a power law, $\rho_e \propto r^{-7}$, and a CSM with a power law in density, $\rho_a \propto r^{-2}$, i.e., $n = 7$ and $s = 2$ in the SSDW model of Chevalier (1982). The CSM also possessed a variation in density with polar angle (from

Blondin & Lundqvist [1993]),

$$\rho(\theta) = C\langle\rho\rangle \left[1 - A\frac{\exp(-2\beta\cos^2\theta) - 1}{\exp(-2\beta) - 1}\right], \tag{1}$$

where $\langle\rho\rangle = \dot{M}/4\pi r^2 v$ is the angle-averaged mass loss rate. The parameter $A$ is the value of the wind asymmetry such that $\alpha = 1/(1 - A)$ is the ratio of density at the equator to density at the pole, and $\beta$ is a steepness parameter. The angular dependence of the mass loss rate given by this function for the values of $\beta$ used in the simulations is shown in Figure 1. For values of $\beta \lesssim 1$ the density varies gradually from pole to equator, while for $\beta \gg 1$ the density is uniform over most of the circumstellar region, but ramps up quickly to the maximum value in a very narrow region near the equator. Thus, large values of $\alpha$ and $\beta$ would resemble a dense circumstellar disk, whereas a more gradual density variation is close to what Asida & Tuchman (1995) find for simulations of rotating red giant variables. The normalization constant $C$ ensures the same total mass loss rate independent of asymmetry values.

## 3. RESULTS

An extensive series of simulations were run on a grid of 340 radial by 128 angular zones with $n = 7$ and $s = 2$. The asymmetry parameters were varied with $\alpha$ ranging from 2 to 16 and $\beta$ ranging from 0.1 to 8. Each simulation was run for at least 20,000 time steps, corresponding to an expansion of over 8 orders of magnitude in radius.

### 3.1. Typical Evolution

In all the simulations that we computed, the dense shell of shocked gas near the back of the interaction region was subject to the Rayleigh-Taylor instability described in detail in Chevalier et al. (1992). This instability reaches saturation at a time of order $100t_o$, or approximately 5 doubling times. Thereafter, the structure of the unstable region remains steady-state in an averaged sense. In the case of no asymmetry, the RT fingers reach a length of approximately 1/2 the width of the interaction region, with relatively little effect on the forward shock. This result is not always the case in the presence of an asymmetry in the ambient wind. In many of our simulations the RT fingers push all the way out to the forward shock, producing a local bulging of the shock front. This effect is attributable to the obliquity of both the forward and reverse shocks. In the case of an asymmetric SSDW, the forward shock is typically propagating into the CSM at an oblique angle, while the supersonic ejecta may be impinging the reverse shock at an oblique angle. The postshock flow behind both shocks is then not purely radial, and the resulting tangential flow may substantially aid the growth of RT fingers. In particular when the forward/reverse shock pair forms an inflection point, such as at the polar tip, or near the cusp seen in high-$\beta$ models, the obliquity of the shocks tends to create a pair of opposing vortices. The updraft (flow in the positive radial direction) created between these paired vortices will enhance the growth of RT fingers (which are normally accompanied by similar, but much weaker pairs of vortices), leading to RT fingers that extend all the way to the forward shock. Once the finger pushes on the forward shock, the bulge of the shock front increases the obliquity of the shock. This enhances the strength of the vortices, leading to a virtually permanent RT finger. This is in contrast to the spherical case where the RT fingers extend out, fall over, and advect back into the thin shell of shocked ejecta.

Figure 2 shows the result of a calculation for a high-$\beta$ wind, so that the wind is spherically symmetric except for a narrow region close to the equatorial plane. There is an indentation in the equatorial plane as expected, but there are also two slight bulges in the forward shock caused by unusually large RT fingers. One of these protrusions lies just above the indentation near the equatorial plane, and is created by the obliquity of the forward and reverse shocks (resulting from the asymmetric CSM) as described above. The other protrusion lies on the polar axis, and although it has the characteristics described above, we believe that it is an artifact of the axial symmetry in the 2-dimensional calculation and would not occur in a 3-dimensional calculation. The reason is that a RT finger on the symmetry axis does not have a chance to bend and fall over, as in other parts of the shell. However, as described in the next subsection, we have found that an angular density gradient on the axis can greatly enhance the growth of a protrusion on the axis. We performed a 3-dimensional calculation (§3.3) to show that this enhanced growth is not simply the result of the symmetry axis.

### 3.2. Dependence on Asymmetry

The asymmetry of the SSDW is a monotonic function of the asymmetry in the circumstellar density distribution. However, the asymmetry of the SSDW is not easy to quantify because it varies significantly with time. In virtually all cases the SSDW does not approach a truly constant shape due to subsonic flow within the interaction region, e.g., the Rayleigh-Taylor instability. To account for the temporal variations in shape while attempting to illustrate the asymmetry in the SSDW, we have plotted the time history of the aspect ratio, $R_p/R_e$, of the SSDWs in Figures 3 and 4. The interior flow leads to many bumps and wiggles in these plots of aspect ratio, but the ratio tends to hover around a well-defined value.

For comparison, we have plotted the aspect ratio expected under the assumption of purely radial flow. This would be expected if the sound speed in the interaction region was much slower than the expansion velocity of the SSDW. Alternatively, one might try to solve for the shape of the forward shock under the assumption of uniform pressure within the interaction region, i.e., in the limit that the sound speed is much larger than the expansion velocity. However, in this problem the expansion velocity is given by the shock velocity of the forward shock, $v_{sh}$, and the sound speed immediately behind a strong adiabatic shock is given by

$$c^2 = \gamma \frac{2(\gamma-1)}{(\gamma+1)^2} v_{sh}^2.$$

Thus the sound speed in the intershock region is always of order the expansion velocity, and neither approximation will be valid. Nonetheless, we would expect that the assumption of purely radial flow will give a lower limit to the aspect ratio. The tangential flow behind the shock will be driven by the high pressure at the equator relative to the pole, creating a flow that will decrease the postshock pressure at the equator, and increase the postshock pressure at the pole, similar to what is seen in the simulations by Blondin & Lundqvist (1993). This in turn will drive the aspect ratio farther from unity. From dimensional analysis, the radius of the SSDW under the assumption of radial flow is $R \propto \rho(\theta)^{1/(s-n)}$ (Chevalier 1982). A lower limit on the aspect ratio is thus $R_p/R_e = \alpha^{1/(n-s)}$. At the other extreme, the assumption of a spatially constant pressure in the interaction region leads to $R \propto \rho(\theta)^{-1/2}$, or $R_p/R_e = \alpha^{1/2}$.

As seen in Figure 3 the aspect ratio of the SSDW increases uniformly with increasing asymmetry. The strong variations in the aspect ratio over time are attributed to the growth and motion of Rayleigh-Taylor fingers emanating from the shell of shocked ejecta. In particular the forward shock is modified by two such fingers: one at the polar axis and one at the cusp formed just above the equator in high-$\beta$ models (see Figure 2 corresponding to the run in Figure 3 with $\alpha = 2$). As the SSDW evolves, the strong vortex flow within the interaction region tends to drag smaller RT fingers toward these two sustained fingers. When these smaller features merge with one of the larger fingers the sustained finger pushes out a little farther on the forward shock, leading to a change in the overall aspect ratio. In particular, the sharp rises in the aspect ratio for $\alpha = 2$ seen in Figure 3 are associated with the mergers of two RT fingers with the RT finger fixed on the polar axis (see Figure 2).

As the value of $\beta$ is decreased the aspect ratio stays relatively constant, until $\beta$ reaches a critical value where the aspect ratio suddenly increases. This effect is illustrated in the series of plots shown in Figure 4. For all values of $\alpha$ used in our simulations, low values of $\beta$ produced aspect ratios far greater than the apparently constant aspect ratio seen at high values of $\beta$. Furthermore, there appears to be only two self-similar values of the aspect ratio for a given value of $\alpha$; a low value (consistent with expectations) for high values of $\beta$, and an anomalously high value for low values of $\beta$ (see, e.g., $\alpha = 8$ in Figure 4). This critical value of $\beta$ below which the aspect ratio is anomalously high is slightly dependent on the value of $\alpha$. For $\alpha = 2$ the SSDW with $\beta = 1$ was similar to the high-$\beta$ simulations with an aspect ratio of about 1.2, while the SSDW with $\beta = 0.1$ produced an aspect ratio of almost 2. In contrast, for $\alpha = 16$ an anomalously large aspect ratio was achieved with a value of $\beta$ as high as 1. Note also that the value of this anomalously high aspect ratio increases monotonically with $\alpha$.

These large aspect ratios in the case of small values of $\beta$ are attributed to the formation of a "protrusion" extending out along the symmetry axis of the circumstellar density distribution, i.e., the SSDW appears to "poke through" the low density hole in the CSM. An example of a SSDW with such large aspect ratio is shown in Figure 5, where we have displayed the results from a model with $\alpha = 4$ and $\beta = 0.1$. This particular run was computed on a grid of 620 radial zones by 320 angular zones in order to better resolve the flow in the narrow protrusion, and to resolve the flow in the interaction region which is now a relatively small fraction of the total radial extent of the grid. Furthermore, the angular zones were unevenly spaced such that roughly half of the zones are within $\pi/8$ of the polar axis. The overall shape and evolution of this model does not differ significantly from a similar model run on our standard low resolution grid.

The growth of this protrusion is similar to the enhanced growth of RT fingers discussed above: the obliquity due to the inflection of the forward and reverse shocks at the pole leads to a pair of vortices (actually a vortex ring around the pole in three dimensions), which aid and sustain the growth of a RT finger on the polar axis. However, in cases with an anomalously large aspect ratio, this "finger" keeps growing far beyond the original spherical shell, forming the observed protrusion. The SSDW eventually reaches a self-similar state and the aspect ratio becomes relatively constant. The protrusion appears to be a local phenomenon and does not depend on the pressure-driven flow of gas from the vicinity of the equator. Even at low $\alpha$, the ratio $R_p/R_e$ can be larger than that expected from a constant pressure shell. The large value of $R_p/R_e$ is driven by the concentration of a relatively large amount of ejecta into a small region of the decelerating flow.

The flow pattern in the protrusion is illustrated in more detail in Figure 6. In this self-similar state the CSM strikes the forward shock of the polar protrusion at a very oblique angle. The shocked CSM is thus directed almost straight down, parallel to the polar axis. The supersonic (in the self-similar frame) downdraft along the outside of the protrusion is decelerated in a strong shock at the base of the protrusion. The high gas pressure between the nominal forward and reverse shocks squeezes the flow toward and up the polar axis. The vortex ring created by this redirection of the flow forms a de Laval nozzle on the axis, which accelerates the flow up the polar axis to Mach numbers of order 5. The updraft is further collimated by a series of weak incident shocks, as seen in the plot of gas pressure in Figure 6. By the time the flow reaches the head of the protrusion it is confined to a very narrow "jet" with a width of only one-tenth the length of the protrusion. At the top of the SN this updraft is decelerated in another strong shock, similar to the working surface associated with supersonic jet outflows.

The parameter determining the presence of a polar protrusion appears to be the gradient in the circumstellar density distribution as a function of angle near the pole. For the circumstellar distribution given by equation (1), the gradient is zero at the pole, so we

consider

$$\frac{d\ln\rho}{d\ln\theta} = -(\alpha - 1)\left(\frac{e^{-2\beta}}{e^{-2\beta} - 1}\right) 4\beta.$$

For the two models that appear just on the edge of an anomalous solution ($\alpha = 4, \beta = 1$ and $\alpha = 2, \beta = 0.1$), this quantity is 1.8. Given that the polar protrusion depends only on the density distribution within $\theta \sim \pi/8$ of the polar axis, we note that the density at the pole need only be $\sim 15\%$ smaller than the rest of the CSM (See Fig. 1). For comparison, the asymmetry function used in Luo & McCray (1991) gives

$$\frac{d\ln\rho}{d\ln\theta} = 2(\alpha - 1).$$

So for $\alpha \gtrsim 2$ the Luo & McCray function should produce SSDW's with an anomalously large aspect ratio. For $\alpha = 2$ our simulation using the Luo & McCray function did indeed produce a protrusion, resulting in a SSDW with an aspect ratio of $\sim 2$.

We note that our results are nearly unaffected by radiative cooling of the shocked ejecta. Using the numerical methods of Chevalier & Blondin (1995) for treating cooling of the shocked ejecta, we computed a simulation with $\alpha = 4$ and $\beta = 1$. These parameters were chosen because they are close to the transition from the expected asymmetry to the anomalous solution. The resulting evolution was was similar to the purely adiabatic case, forming a protrusion only after extended evolution (see Figure 4 for the adiabatic case).

### 3.3. Three-Dimensional Simulation

A significant limitation of the two-dimensional simulations described above is the assumption of axisymmetry. This assumption is particularly questionable in the low-$\beta$ cases when a protrusion extends up the symmetry axis. The requirement of reflection symmetry on the axis means that this protrusion cannot be deflected off to one side or the other; it will just continue to grow along the axis. To check the dependence of the anomalous aspect ratio on the forced axisymmetry in the two-dimensional simulation, we ran a three-dimensional simulation with $\alpha = 10$ and $\beta = 1$.

The 3D simulation was run on a numerical grid of 235 ($r$) by 128 ($\theta$) by 128 ($\phi$) zones. The grid was laid out with the polar axis of the spherical grid in the equatorial plane of the circumstellar material so that the symmetry axis of the circumstellar density distribution did not coincide with any unique axis on the numerical grid. Thus $\theta$ and $\phi$ both ranged from 0 to $\pi$. Note that this leads to a spatial resolution in the angular direction of only half that used in the 2D models. To avoid severe constraints on the time step as a result of vanishingly small zone sizes in the $\phi$ direction near the polar axis of the grid, the density, pressure and $\phi$ velocities were smoothed over within a small cone about the axis. While this produces a noticeable effect in the simulation, it only affects a small region in the equatorial plane of the SSDW and therefore does not affect the goal of the simulation: to quantify the aspect ratio of the SSDW and the stability of the polar protrusion.

The 3D model evolved almost identically to the corresponding 2D simulation. Figure 7 shows a slice of the 3D model next to a plot from a 2D simulation with identical parameters and at roughly the same evolutionary time. Note that the SSDW has not yet reached the steady-state value of the aspect ratio, which for this case should be $\sim 3.8$. The shape and flow pattern in the two simulations appears qualitatively similar. In particular, the 3D model has a similar polar protrusion to that seen in the 2D models. The flow of material up the polar axis is not entirely straight due to the influence of the Kelvin-Helmholtz instability expected in the presence of the strong shear flow between the down- and updrafts within the protrusion. Although the flow is not exactly axisymmetric, it appears that the assumption of axisymmetry is reasonable.

To test the stability of the vortex tube and associated polar protrusion, we evolved the 3D SSDW into an asymmetric circumstellar medium. A cloud with 3 times the ambient density was placed on one side of the blastwave in an attempt to "knock" the protrusion over. This experiment failed to produce any significant effects, suggesting that this unusual flow pattern is relatively robust.

### 3.4. Dependence on Ejecta Density Profile

The effect of an axisymmetric CSM is expected to be more pronounced for smaller values of $n$ because of the stronger dependence of the expansion velocity on the circumstellar density in such cases (Chevalier 1982). To examine the effect of varying $n$ we have run simulations with steeper ejecta density profiles: $n = 10$ and 15.

For simulations with large $\beta$, where the anomalous aspect ratio does not come into play, the aspect ratio behaves as expected. Figure 8 shows the evolution of the aspect ratio for runs with $\alpha = 4$ and $\beta = 4$. At late times the aspect ratio is comparable to the analytic estimate of $\alpha^{1/n-s}$.

Given the smaller aspect ratio of higher-$n$ SSDW's, one would also expect the need for a larger density gradient in order to stimulate the growth of a polar protrusion. Figure 8 shows the aspect ratio for simulations with $\beta = 0.1$, which in the case of $n = 7$ produced an anomalous aspect ratio of $\sim 2.5$. For $n = 10$ a protrusion does begin to form at the pole, but it is much smaller than the $n = 7$ SSDW. For $n = 15$ there is a RT finger that pushes on the forward shock at the pole, but it is not able to generate a large protrusion as with lower-$n$ simulations. If, on the other hand, the asymmetry in the CSM is increased to produce the same canonical aspect ratio in an $n = 10$ SSDW as in a SSDW with $\alpha = 4$ and $n = 7$ (i.e., $\alpha = 10$ for $n = 10$), the polar protrusion is again unusually large, and is in fact very similar to that in the $n = 7$ simulation. The critical density gradient in the CSM needed to produce a polar protrusion is thus larger for larger values of the ejecta exponent, $n$.

### 3.5. Axisymmetric Ejecta

Finally, we consider the complementary problem of axisymmetric stellar ejecta driving a SSDW into a spherically symmetric CSM. We use the same asymmetry function (eqn. [1]) to describe the density distribution of the stellar ejecta, and evolve the SSDW into a spherically symmetric CSM. The result is a much smaller effect on the aspect ratio of the SSDW. For $\alpha = 4$ and $\beta = 0.1$ the aspect ratio of the remnant is only $\sim 0.9$, with the equatorial radius larger than the polar radius. The assumption of purely radial flow predicts an aspect ratio of 0.76.

This difference in the resulting aspect ratio can be attributed to at least two effects: changes in the pressure of the intershock region due to changes in shock radius, and the direction of tangential flow. In the case of an asymmetric CSM, the higher density CSM pinches in the SSDW, which tends to increase the postshock pressure. Any tangential flow generated by this increased pressure will be away from the high density CSM and hence will increase the aspect ratio. In the case of asymmetric ejecta, the higher density ejecta pushes out on the SSDW. While the higher density ejecta will increase the pressure of the intershock region, pushing the shock outwards will lower this pressure (as opposed to the asymmetric CSM, which increases the pressure when it moves the shock inwards). Furthermore, in the asymmetric ejecta case, any tangential flow generated by the overpressure region will serve to drive the aspect ratio back towards unity.

## 4. DISCUSSION

The most novel aspect of our calculations is the development of protrusions for moderate values of the angular density gradient at the poles. Only a mild asymmetry ($\rho_e/\rho_p \sim 2$) is needed to form the protrusion, as long as there is a moderate angular density gradient at the pole. Possibly related protrusions have been found in models of the interacting winds scenario for planetary nebulae and other objects. Icke et al. (1992) found the growth of a jet-like feature along the polar axis if they set up a low density in the slow external wind along the polar axis. The angular dependence of density that they chose in the slow wind is the inverse of that used here andtends to create a channel along the polar axis. In both cases, the backflow down the sides of the narrow outflowing gas helps to keep the protrusion in place. In the planetary nebula case, the collimation is aided by the fact that the termination shock of the fast, central wind is elongated in the polar direction so that the newly shocked gas receives a velocity component directed toward the pole. This effect does not appear to be especially significant in our calculations (see the reverse shock front in Figs. 5 and 6). What is significant is the presence of a reverse shock relatively close to the forward shock such that the backflow is confined and redirected back up the polar axis. In addition, the deceleration of the interaction shell leads to Rayleigh-Taylor instabilities and vortical motion in the shell that aid in the formation of the protrusion. In essence, the vortex surrounding the polar axis becomes highly elongated.

The clearest application of the calculations of protrusion development is to the remnant 41.9 +58 in M82. The radio image shows a compact shell (so that circumstellar interaction is plausible) with two oppositely directed protrusions (Bartel et al. 1987; Wilkinson & deBruyn 1990). As noted by these authors, the morphology is unusual for a SN remnant. We would identify the long axis with the symmetry axis of an axisymmetric dense presupernova wind. A remaining problem is whether there is sufficient time for the protrusions to develop because Fig. 4 shows that a large increase in time can be necessary to approach the self-similar state. The presence of a dense wind implies that the SN progenitor was red supergiant star. If the initial stellar radius is $3 \times 10^{13}$ cm and the shock velocity is 10,000 km s$^{-1}$, the doubling time for the radius is $3 \times 10^4$ s. We identify this time with $t_o$. Figure 4 shows that the $\alpha = 8$, $\beta = 0.1$ run reaches an aspect ratio of 2.5 by $t/t_o = 10^4$, or 10 years for the SN parameters. The estimated age of 41.9 +58 at the time of the VLBI observations is 30 yr, so there is sufficient time for the development of the protrusions. These estimates show that observations covering the initial years of evolution of a SN may show the development of protrusions, although fig. 4 shows that observations covering at least an order of magnitude in age are needed to show significant growth.

The fact that of order 10 years is needed to form a protrusion means that the progenitor must have experienced persistent asymmetric mass loss towards the end of the RSG stage. A minimum time is $\sim 10^4$ years for a wind velocity of 10 km s$^{-1}$. This may be long for a common envelope phase, but short compared to the typical duration of the RSG stage, $\sim 10^6$ years. Any mechanism causing circumstellar asymmetry and which operated for the last $\sim 1\%$ of the RSG phase would probably be sufficient in the case of 41.9 +58.

The other SN with asymmetric VLBI structure, SN 1986J, is a less promising case for comparison with our theory because the image shows 3 protrusions at roughly 120° intervals (Bartel et al. 1991). This is not compatible with the axisymmetric structure that we have assumed here. However, there is evidence that the CSM around SN 1986J is inhomogeneous (Chugai & Danziger 1994) and there is the possibility that inhomogeneities are able to trigger the growth of protrusions.

This research was supported in part by NASA grant NAG 5-2844, NSF grant AST-9314724, and the Swedish Natural Sciences Research Council. The numerical simulations were computed on a Cray YMP at the North Carolina Supercomputing Center.

# REFERENCES

Bartel, N., et al. 1987, ApJ, 323, 505

Bartel, N., Rupen, M. P., Shapiro, I. I., Preston, R. A., & Rius, A. 1991, Nature, 350, 212

Blondin, J. M. 1994, in Circumstellar Media in the Lates Stages of Stellar evolution, ed. R. E. S. Clegg, W. P. S. Meikle, & I. R. Stevens (CUP:Cambridge), 139

Blondin, J. M., & Lundqvist, P. 1993, ApJ, 405, 337

Bowers, P. F., Johnston, K. J., & De Vegt, C. 1989, 340, 479

Burrows, C. J., et al. 1995, ApJ, 452, 680

Chevalier, R. A. 1982, ApJ, 258, 790

Chevalier, R. A. 1984, Ann. N. Y. Acad. Sci., 422, 215

Chevalier, R. A., & Blondin, J. M. 1995, ApJ, 444, 312

Chevalier, R. A., Blondin, J. M., & Emmering, R. T. 1992, ApJ, 392, 118

Chevalier, R. A., & Dwarkadas, V. V. 1995, 452, L45

Chevalier, R. A., & Fransson, C. 1994, ApJ, 420, 268

Chevalier, R. A., & Luo, D. 1994, ApJ, 421, 225

Chugai, N. N., & Danziger, I. J. 1994, MNRAS, 268, 173

Colella, P., & Woodward, P. R. 1984, J. Comp. Phys., 54, 174

Crotts, A. P. S., Kunkel, W. E., & Heathcote, S. R. 1995, ApJ, 438, 724

Cumming, R., Lundqvist, P., Sollerman, J., & Meikle, W. P. S. 1996, in preparation

Fesen, R. A., Martin, C. L., & Shull, J. M. 1992, ApJ, 399, 599

Icke, V., Mellema, G., Balick, B., Eulderink, F., & Frank, A. 1992, Nature, 355, 524

Igumenshchev, I. V, Tutukov, A. V., & Shustov, B. M. 1992, Sov.Astron., 36, 241

– 20 –

Jakobsen, P. et al. 1991, ApJ, 369, L63

Kahn, F. D., & West, K. A. 1985, MNRAS, 212, 837

Kwok, S., Purton, C. R., & Fitzgerald, P. M. 1978, ApJ, 219, L125

Luo, D., & McCray, R. 1991, ApJ, 379, 659

Luo, D., McCray, R., & Slavin, J. 1994, ApJ, 430, 264

Marcaide, J. M. 1995a, Nature, 373, 44

Marcaide, J. M. 1995b, Science, 270, 1475

Martin, C. L., & Arnett, D. A. 1995, ApJ, 447, 378

Mauron, N., & Querci, C. 1990, A&AS, 86, 513

McCray, R., & Lin, D. N. C. 1994, Nature, 369, 378

Plait, P. C., Lundqvist, P., Chevalier, R. A., & Kirshner, R. P. 1995, ApJ, 439, 730

Plez, B., & Lambert, D. L. 1994, ApJ, 425, L101

Podsiadlowski, Ph., Fabian, A. C. & Stevens, I. R. 1991, Nature, 354, 43

Seward, F. D. 1990, ApJS, 73, 781

Trammel, S. R., Dinerstein, H. L., & Goodrich, R. W. 1994, AJ, 108, 984

Wampler, E. J., Wang, L., Baade, D., Banse, K., D'Odorico, S., Gouiffes, C., & Tarenghi, M. 1990, ApJ, 362, L13

Wang, L., & Mazzali, P. A. 1992, Nature, 355, 58

Wang, L., & Wampler, E. J. 1992, A&A, 262, L9

Wilkinson, P. N., & de Bruyn, A. G. 1990, MNRAS, 242, 529

Zuckerman, B., & Aller, L. H. 1986, ApJ, 301, 772

Fig. 1.— Function used to describe the asymmetry of the circumstellar matter in the hydrodynamic simulations for different values of $\beta$ assuming $\alpha = 4$.

Fig. 2.— Density (right) and Velocity (left) of an evolved SSDW propagating into an asymmetric CSM described by $\beta = 8$, $\alpha = 2$. The density in the right image is displayed with both logarithmic contours and shading, although the shading is only used within the interaction region to highlight the shell of dense shocked ejecta and the Rayleigh-Taylor fingers that emanate from this shell. The velocity vectors in the left image are plotted in the expanding self-similar frame, and the locations of the forward and reverse shocks are marked with heavy solid lines. No vectors are plotted for the undisturbed CSM or ejecta.

Fig. 3.— Evolution of the aspect ratio ($R_p/R_e$) of a SSDW propagating into an asymmetric CSM described by a value of $\beta = 8$ and a value of $\alpha$ as labelled. The dashed lines are the values of the aspect ratio predicted in the limit of no tangential flow for the same values of $\alpha$.

Fig. 4.— Evolution of the aspect ratio of an a SSDW propagating into an asymmetric CSM. Each plot shows the evolution of four simulations corresponding to $\beta = 0.1$, 1, 4, and 8 ($\alpha = 16$ does not have a curve for $\beta = 0.1$). In each case the simulation with the largest aspect ratio corresponds to the smallest value of $\beta$.

Fig. 5.— A protrusion along the polar axis is evident in simulations with low values of $\beta$, as in this model with $\beta = 0.1$ and $\alpha = 4$. The right-hand image displays the density as in Figure 2. The left-hand image displays the entropy of the gas, with dark shading representing low entropy gas corresponding to the shocked ejecta. This simulation was computed on a high-resolution grid with 600 radial and 320 angular zones.

Fig. 6.— A close up of the polar protrusion from the model shown in Fig. 5. The shading on the right corresponds to the logarithm of the gas pressure, spanning a range of a factor of 2 in the logarithm. For reference, the velocity vectors are plotted every fourth numerical zone in the angular direction, and every eighth zone in the radial direction.

Fig. 7.— A two-dimensional slice of the 3D model (left) appears qualitatively similar to an axisymmetric 2D model (right) with the same parameters (in this case $\beta = 1$, $\alpha = 10$). Note that these simulations have not yet reached the maximum aspect ratio expected for these parameters.

Fig. 8.— Aspect ratio of SSDW as a function of the power-law exponent of the ejecta density profile, $n$. For large values of $\beta$ (*left*) the aspect ratio decreases with increasing $n$, in agreement with analytic prediction (marked by the dashed lines). For small values of $\beta$ (*right*) the polar protrusion is prominent for low $n$, present, but small for $n = 10$, and essentially absent for $n = 15$.